\documentclass[letterpaper]{article}
\usepackage{aaai}
\usepackage{times}
\usepackage{helvet}
\usepackage{courier}
\usepackage{multirow}
\usepackage{graphicx}
\usepackage{caption}
\usepackage{subcaption}
\usepackage{color,hyperref}
    \catcode`\_=11\relax
    
    \def\_email#1@#2\q_nil{%
      \href{mailto:#1@#2}{{\emailfont #1\emailampersat #2}}
    }
    \newcommand\emailfont{\sffamily}
    \newcommand\emailampersat{{\color{red}\small@}}
    \catcode`\_=8\relax 
\begin{document}
\title{Two-step adversarial debiasing with partial learning - medical image case-studies}
\author{Ramon Correa$^{1}$, Jiwoong Jason Jeong$^{1}$, Bhavik Patel$^{2,1}$, Hari Trivedi$^{3}$, Judy W. Gichoya$^{3}$, Imon Banerjee$^{2, 1}$ \\
$^{1}$SCAI, Arizona State University, USA \\ 
$^{2}$Department of Radiology, Mayo Clinic, Arizona, USA \\
$^{3}$Department of Radiology, Emory University, Atlanta, USA \\
\href{mailto:imon.banerjee@asu.edu}{imon.banerjee@asu.edu}
}
\maketitle
\begin{abstract}
\begin{quote}
The use of artificial intelligence (AI) in healthcare has become a very active research area in the last few years. While significant progress has been made in image classification tasks, only a few AI methods are actually being deployed in hospitals. A major hurdle in actively using clinical AI models currently is the trustworthiness of these models. More often than not, these complex models are black boxes in which promising results are generated. However, when scrutinized, these models begin to reveal implicit biases during the decision making, such as detecting race and having bias towards ethnic groups and subpopulations. In our ongoing study, we develop a two-step adversarial debiasing approach with partial learning that can reduce the racial disparity while preserving the performance of the targeted task. The methodology has been evaluated on two independent medical image case-studies - chest X-ray and mammograms, and showed promises in bias reduction while preserving the targeted performance. 
\end{quote}
\end{abstract}

\section{Introduction}

%use these papers as reference: https://aaai.org/Library/AAAI/aaai20contents-issue07.php#21
Artificial Intelligence (AI) models have demonstrated expert-level
performance in image-based diagnostic tasks, resulting in increased clinical adoption
and FDA approvals. The new challenge in AI is to understand the limitations of
models from the perspective of demographic bias in order to reduce potential harm. The unknown disparities based on demographic factors could worsen currently existing inequalities worsening patient care for some groups.

AI bias can be defined as models with outputs providing outcomes that negatively affects one sub-group of the study population more than others. Examples include differing allocation of healthcare resources based on patient demographics~\cite{racialBiasHealth,riskAutomatingRacism}, bias in language models developed on clinical notes~\cite{HurtfulWordsQuantifyingBias}, and melanoma detection models developed primarily on images of light-colored skin \cite{dermatologyFairness}. In the clinical domain, unintended bias in AI systems affecting individuals unfairly based on race, gender, and other clinical characteristics has been highlighted in multiple studies~\cite{addressingBiasinAI,disabilityBiasandAi}. AI models applied in other applications have also presented similar biases such as face detection models failing to correctly identify individuals of minority groups \cite{GenderShades}. Given such examples, biased AI systems can result in a variety of fairness-related harms, particularly in healthcare.

A core challenge for reducing AI bias is that the model final performance and reasons resulting in unfairness of AI models are not mutually exclusive and can often exacerbate one another. Recently, \cite{banerjee2021reading} showed that AI models trained for diagnosis can learn unintended racial information from different imaging modalities. Thus, AI models may use learned demographic information for detecting a diagnosis even when such attribute is not associated with the diagnosis. There are examples of race-ethnicity and gender influencing clinician decision-making, and given that AI is trained on real-world data, it is reasonable to expect that models would learn similar biases. 

Common technique adopted within the community to reduce biases is curating a training dataset with greater number of positive cases across demographics~\cite{larrazabal_gender_2020}. Other popular approaches to remove biases such as building demographic-specific models often suffer from a lack of demographic representation. We observed that  techniques attempting to decouple demographic information and task predictions are not able to match baseline model performances~\cite{seyyed-kalantari_chexclusion_2020}. Our goal is to develop an efficient methodology for model debiasing without the need for demographically balanced datasets and simultaneously match the baseline model performance.

The core contributions of the current article are - 
\begin{enumerate}
\item Present AI model bias in terms of patients' race for two prevailing use-cases - chest X-ray and mammogram image interpretation. 
\item Reducing racial bias by implementing a novel adversarial debiasing technique with partial model tuning while preserving the baseline model performance.
\item Compare the performance of full and partial debiasing techniques for both chest X-ray and mammogram image interpretation.  
\end{enumerate}

\section{Methodology}
\begin{figure}[h!]

\includegraphics[width=0.5\textwidth]{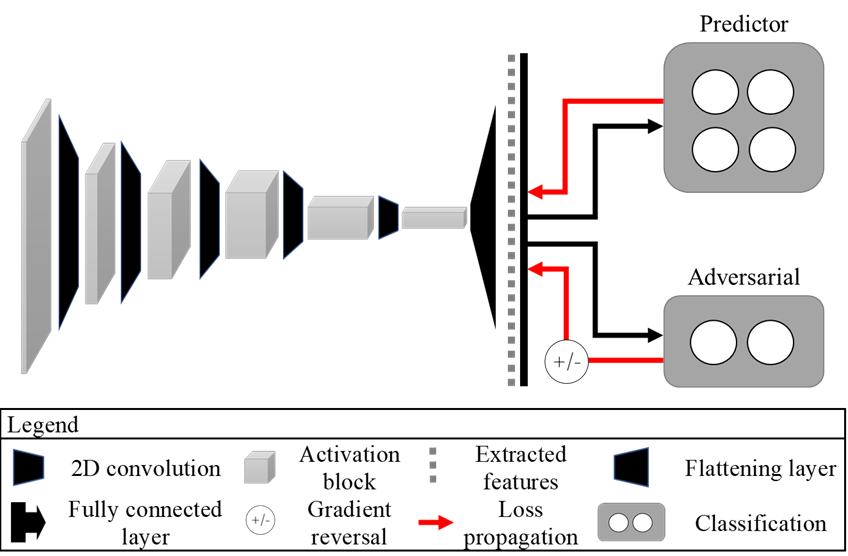}
  \caption{Architecture of the proposed architecture - adversarial and predictor  branch\label{fig:comparative_inference}}
\centering
\end{figure}

In Fig.~\ref{fig:comparative_inference}, we present a simplistic visual of the proposed architecture which contains two parallel branches after core CNN backbone network -  (1) \textbf{predictor} - train to predict targeted variable $y$ given input $X$ by minimizing cost $L_{predictor}(y, \hat{y})$. $\hat{y}$ is the model prediction given input $X$ which can be modeled as a $f(X)$; (2) \textbf{adversarial} - predict the protected variable $Z$ given input $X$ and reverse the gradient for penalizing learning of protected variable. Hypothetically, often $f(X)$ is highly dependent on protected variable $Z$ and penalizing the learning of protected often significantly hamper the prediction performance of the target data. Demographic factors, including race, can be modeled as a protected variable. 

Training of the proposed architecture involves two backward passes. In the first pass, the model minimize the loss for both predictor and adversarial branches as $L=L_{predictor}+ \lambda *L_{adversarial}$ where $L_{adversarial}$ is loss for learning the protected variable $Z$ and $\lambda$ is the regularization factor which can be in the range of $[0,1]$. In the second pass, we only backpropagate the sign flipped gradient corresponding to the adversarial branch for modeling the penalization as $L=-\lambda*L_{adversarial}$. First pass helps the model to learn simultaneously the targeted and protected variables while the second pass intend to suppress learning of rotected variable $Z$. We used the same $\lambda = 0.53$ for both phases.    

\noindent\textit{Partial fine-tuning}: We explore the potential of improving model performance even after de-biasing by fine-tuning a subset of convolution layers of a pre-trained model instead of the complete network. The layers used for finetuning are identified via an \textit{ablation study} which measures the performance difference of the targeted and protected variable prediction after dropping 10\% of the top similar filters relative to the total number of filters from the targeted convolution layers~\cite{meyes2019ablation}. It should be noted that due to the increasing sizes of the different convolutional layers, the same proportion may correspond to a different number of ablated filters. Layers where the protected variable prediction experienced higher performance degradation, was identified for finetuning alongside its downstream layers with the proposed two-step adversarial learning.  

\section{Results}
%(CXR Lambda: .40, mammo_lambda:.53)
\subsection{Datasets}

In this study, we validated the performance of the proposed architecture for the following two independent use-cases - 
\begin{table}[]
\centering
\caption{Dataset description - chest X-ray and mammography images. Tissue density 1 = fatty, 2 = fibrogranular density, 3 = heterogeneously dense, 4 = extremely dense.}
\label{tab:demographics}
\resizebox{0.4\textwidth}{!}{%
\begin{tabular}{cccccc}
\hline
\multicolumn{6}{|c|}{\textit{Chest   X-ray Dataset}}      \\ \hline
\multicolumn{2}{|c|}{Race}                                                                                                                                & \multicolumn{2}{c|}{Gender}                                                                 & \multicolumn{2}{c|}{Age}                                                                   \\ \hline
\multicolumn{1}{|c|}{\multirow{5}{*}{\begin{tabular}[c]{@{}c@{}}Black/\\ African\\ American\end{tabular}}} & \multicolumn{1}{c|}{\multirow{5}{*}{38,024}} & \multicolumn{1}{c|}{\multirow{5}{*}{Male}}   & \multicolumn{1}{c|}{\multirow{5}{*}{34,857}} & \multicolumn{1}{c|}{\multirow{2}{*}{0-19}}  & \multicolumn{1}{c|}{\multirow{2}{*}{1,566}}  \\
\multicolumn{1}{|c|}{}                                                                                     & \multicolumn{1}{c|}{}                        & \multicolumn{1}{c|}{}                        & \multicolumn{1}{c|}{}                        & \multicolumn{1}{c|}{}                       & \multicolumn{1}{c|}{}                        \\ \cline{5-6} 
\multicolumn{1}{|c|}{}                                                                                     & \multicolumn{1}{c|}{}                        & \multicolumn{1}{c|}{}                        & \multicolumn{1}{c|}{}                        & \multicolumn{1}{c|}{\multirow{2}{*}{20-39}} & \multicolumn{1}{c|}{\multirow{2}{*}{13,237}} \\
\multicolumn{1}{|c|}{}                                                                                     & \multicolumn{1}{c|}{}                        & \multicolumn{1}{c|}{}                        & \multicolumn{1}{c|}{}                        & \multicolumn{1}{c|}{}                       & \multicolumn{1}{c|}{}                        \\ \cline{5-6} 
\multicolumn{1}{|c|}{}                                                                                     & \multicolumn{1}{c|}{}                        & \multicolumn{1}{c|}{}                        & \multicolumn{1}{c|}{}                        & \multicolumn{1}{c|}{\multirow{2}{*}{40-59}} & \multicolumn{1}{c|}{\multirow{2}{*}{21,227}} \\ \cline{1-4}
\multicolumn{1}{|c|}{\multirow{5}{*}{\begin{tabular}[c]{@{}c@{}}White/\\ Caucasian\end{tabular}}}          & \multicolumn{1}{c|}{\multirow{5}{*}{35,348}} & \multicolumn{1}{c|}{\multirow{5}{*}{Female}} & \multicolumn{1}{c|}{\multirow{5}{*}{38,515}} & \multicolumn{1}{c|}{}                       & \multicolumn{1}{c|}{}                        \\ \cline{5-6} 
\multicolumn{1}{|c|}{}                                                                                     & \multicolumn{1}{c|}{}                        & \multicolumn{1}{c|}{}                        & \multicolumn{1}{c|}{}                        & \multicolumn{1}{c|}{\multirow{2}{*}{60-79}} & \multicolumn{1}{c|}{\multirow{2}{*}{28,374}} \\
\multicolumn{1}{|c|}{}                                                                                     & \multicolumn{1}{c|}{}                        & \multicolumn{1}{c|}{}                        & \multicolumn{1}{c|}{}                        & \multicolumn{1}{c|}{}                       & \multicolumn{1}{c|}{}                        \\ \cline{5-6} 
\multicolumn{1}{|c|}{}                                                                                     & \multicolumn{1}{c|}{}                        & \multicolumn{1}{c|}{}                        & \multicolumn{1}{c|}{}                        & \multicolumn{1}{c|}{\multirow{2}{*}{80+}}   & \multicolumn{1}{c|}{\multirow{2}{*}{8,968}}  \\
\multicolumn{1}{|c|}{}                                                                                     & \multicolumn{1}{c|}{}                        & \multicolumn{1}{c|}{}                        & \multicolumn{1}{c|}{}                        & \multicolumn{1}{c|}{}                       & \multicolumn{1}{c|}{}                        \\ \hline
\multicolumn{2}{|c|}{Total Patients}                                                                                                                      & \multicolumn{4}{c|}{73,372}                                                                                                                                                              \\ \hline
\multicolumn{2}{|c|}{Total Images}                                                                                                                        & \multicolumn{4}{c|}{137,985}                                                                                                                                                             \\ \hline
  
\multicolumn{6}{|c|}{\textit{Mammography Dataset}}                                                                                                                                                                                                                                                                                                            \\ \hline
\multicolumn{2}{|c|}{Race}                                                                                                                                & \multicolumn{2}{c|}{Tissue Density}                                                         & \multicolumn{2}{c|}{Age}                                                                   \\ \hline
\multicolumn{1}{|c|}{\multirow{4}{*}{Asian}}                                                               & \multicolumn{1}{c|}{\multirow{4}{*}{1,305}}  & \multicolumn{1}{c|}{\multirow{3}{*}{1}}      & \multicolumn{1}{c|}{\multirow{3}{*}{1,853}}  & \multicolumn{1}{c|}{\multirow{3}{*}{$<$45}}   & \multicolumn{1}{c|}{\multirow{3}{*}{2,941}}  \\
\multicolumn{1}{|c|}{}                                                                                     & \multicolumn{1}{c|}{}                        & \multicolumn{1}{c|}{}                        & \multicolumn{1}{c|}{}                        & \multicolumn{1}{c|}{}                       & \multicolumn{1}{c|}{}                        \\
\multicolumn{1}{|c|}{}                                                                                     & \multicolumn{1}{c|}{}                        & \multicolumn{1}{c|}{}                        & \multicolumn{1}{c|}{}                        & \multicolumn{1}{c|}{}                       & \multicolumn{1}{c|}{}                        \\ \cline{3-6} 
\multicolumn{1}{|c|}{}                                                                                     & \multicolumn{1}{c|}{}                        & \multicolumn{1}{c|}{\multirow{3}{*}{2}}      & \multicolumn{1}{c|}{\multirow{3}{*}{7,408}}  & \multicolumn{1}{c|}{\multirow{3}{*}{45-59}} & \multicolumn{1}{c|}{\multirow{3}{*}{7,279}}  \\ \cline{1-2}
\multicolumn{1}{|c|}{\multirow{4}{*}{\begin{tabular}[c]{@{}c@{}}Black/\\ African\\ American\end{tabular}}} & \multicolumn{1}{c|}{\multirow{4}{*}{8,164}}  & \multicolumn{1}{c|}{}                        & \multicolumn{1}{c|}{}                        & \multicolumn{1}{c|}{}                       & \multicolumn{1}{c|}{}                        \\
\multicolumn{1}{|c|}{}                                                                                     & \multicolumn{1}{c|}{}                        & \multicolumn{1}{c|}{}                        & \multicolumn{1}{c|}{}                        & \multicolumn{1}{c|}{}                       & \multicolumn{1}{c|}{}                        \\ \cline{3-6} 
\multicolumn{1}{|c|}{}                                                                                     & \multicolumn{1}{c|}{}                        & \multicolumn{1}{c|}{\multirow{3}{*}{3}}      & \multicolumn{1}{c|}{\multirow{3}{*}{7,205}}  & \multicolumn{1}{c|}{\multirow{3}{*}{60-79}} & \multicolumn{1}{c|}{\multirow{3}{*}{6,705}}  \\
\multicolumn{1}{|c|}{}                                                                                     & \multicolumn{1}{c|}{}                        & \multicolumn{1}{c|}{}                        & \multicolumn{1}{c|}{}                        & \multicolumn{1}{c|}{}                       & \multicolumn{1}{c|}{}                        \\ \cline{1-2}
\multicolumn{1}{|c|}{\multirow{4}{*}{\begin{tabular}[c]{@{}c@{}}White/\\ Caucasian\end{tabular}}}          & \multicolumn{1}{c|}{\multirow{4}{*}{7,924}}  & \multicolumn{1}{c|}{}                        & \multicolumn{1}{c|}{}                        & \multicolumn{1}{c|}{}                       & \multicolumn{1}{c|}{}                        \\ \cline{3-6} 
\multicolumn{1}{|c|}{}                                                                                     & \multicolumn{1}{c|}{}                        & \multicolumn{1}{c|}{\multirow{3}{*}{4}}      & \multicolumn{1}{c|}{\multirow{3}{*}{873}}    & \multicolumn{1}{c|}{\multirow{3}{*}{80+}}   & \multicolumn{1}{c|}{\multirow{3}{*}{468}}    \\
\multicolumn{1}{|c|}{}                                                                                     & \multicolumn{1}{c|}{}                        & \multicolumn{1}{c|}{}                        & \multicolumn{1}{c|}{}                        & \multicolumn{1}{c|}{}                       & \multicolumn{1}{c|}{}                        \\
\multicolumn{1}{|c|}{}                                                                                     & \multicolumn{1}{c|}{}                        & \multicolumn{1}{c|}{}                        & \multicolumn{1}{c|}{}                        & \multicolumn{1}{c|}{}                       & \multicolumn{1}{c|}{}                        \\ \hline
\multicolumn{2}{|c|}{Total Patients}                                                                                                                      & \multicolumn{4}{c|}{17,393}                                                                                                                                                              \\ \hline
\multicolumn{2}{|c|}{Total Images}                                                                                                                        & \multicolumn{4}{c|}{34,134}                                                                                                                                                              \\ \hline
\end{tabular}%
}
\end{table}
\noindent(1) \textit{Diagnosis from chest X-ray images} - We received the de-identified dataset of 137,985 chest x-ray images of 73,372 unique patients from Emory University hospital. The demographic factors are described in Table~\ref{tab:demographics}. The targeted task is to identify four common radiology findings - i) atelectasis, 2) edema, 3) pneumothorax, and 4) normal cases, and patient race is considered as protected variable.   

\noindent(2) \textit{Infer tissue density from the mammogram images} - We received the de-identified dataset of 34,134 mammogram images of 17,393 unique patients from Emory University hospital (see Table~\ref{tab:demographics}). The targeted task is to classify the images based on breast tissue density inferred manually by the expert radiologist, and similar to the chest x-ray, patient race is considered as protected variable.
\subsection{Quantitative Performance}
We have compared the performance three models - (1) \textit{Baseline} -  a CNN backbone but with no de-biasing, (2) \textit{Full debias} - 2-step adversarial training of the same CNN backbone, (3) \textit{Partial debias} - 2-step adversarial training of the same CNN backbone with ablation for indentifying optimal layers for learning. In order to evaluate the models, we generated a patient-wise separation of the $train:validation:test$ splits for both case-studies as $60:20:20$ and presented the results of all the models on the test set. 
Table~\ref{tab:results} presents class-wise performance of all the three models in terms of AUCROC, Precision and recall. Bias of the models is evaluated in terms of True Positive Disparity (TPR) score where the true positive rate of the target patient subgroup is compared with the reference group as $\frac{TPR_{target}}{TPR_{ref}}$. any disparity measure between 0.8 and 1.25 can be deemed fair as per the 80 percent rule for determining disparate impact~\cite{corbett2018measure}.  
\begin{figure}
    \centering
    \begin{subfigure}[t]{.35\textwidth}
        \centering
        \includegraphics[scale=.4]{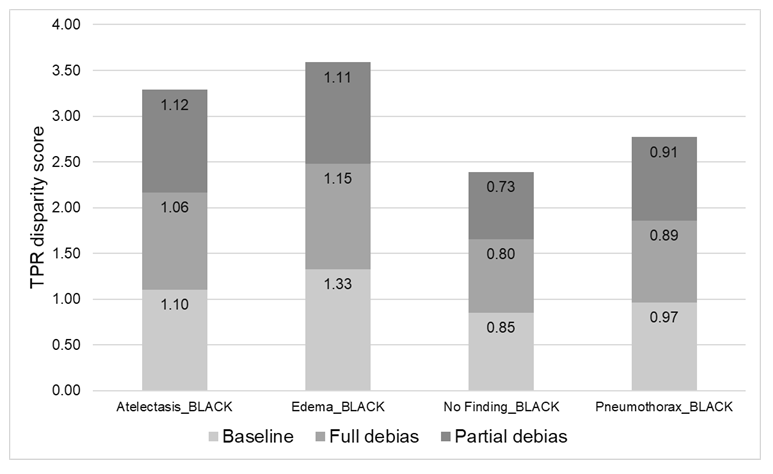}
    \end{subfigure}
    \hfill
    \begin{subfigure}[t]{.35\textwidth}
        \centering
        \includegraphics[scale=.35]{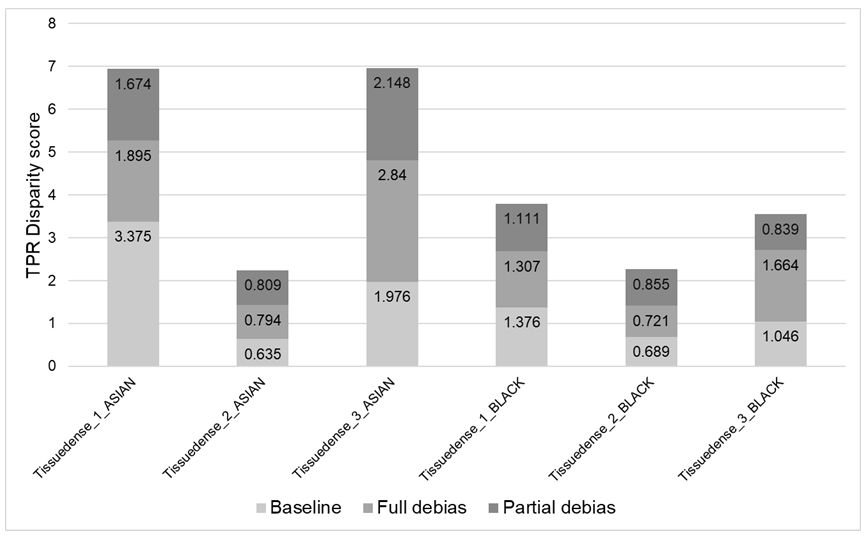}
    \end{subfigure}
   
    \caption{Class-wise TPR disparity plots for baseline, partial and full debias model: top plot presents chest x-ray and bottom plot presents mammogram use case. Caucasian(white) is represented as ref group in both disparity measures. \label{fig:disparity}}
\end{figure}

As seen from Table~\ref{tab:results}, with partial and full debiasing, we achieved comparable overall target task performance on the mammogram images with the baseline (no debiasing). Interestingly, for chest X-ray case-study, we even observed slight performance improvement with partial debiasing over the baseline - no finding AUC improved from 0.86 to 0.89, atelectasis improved from 0.86 to 0.87.  

Chest X-ray cases study is a classic case of low bias AI model where the TPR disparity of the chest X-ray case studies was within the acceptable range for the baseline model, except for the Edema class (see Fig.~\ref{fig:disparity}). With partial debiasing, we managed to reduce the TPR disparity of Edema from 1.33 to the acceptable range of 1.11 with no significant change in performance. For the mammography use-case, the TPR disparity was prominent in baseline model for both African American and Asian patients for all the tissue density classes. Given the known correlation between patient race and breast tissue density, full debias model performance degraded from the baseline while disparity didn't improve. With the new partial de-biasing technique, we preserve the baseline performance while reducing the TPR disparity for African American patients. TPR disparities for Asian patients on the low tissue density classes were improved from 3.37 to 1.89 with full debiasing and to 1.67 with partial debiasing. However, no significant change was observed  in disparity score for heterogeneously dense tissue class for Asian patients. This could be due to the fact that most of the Asian women have denser breasts on mammography thus reducing disparity for denser tissue class is extremely challenging~\cite{bae2016breast}.
% \usepackage{multirow}
% \usepackage{graphicx}

% \usepackage{multirow}
% \usepackage{graphicx}
\begin{table}[]
\centering
\caption{Performance analysis for the targeted task - chest X-ray diagnosis and breast density classification is represented within the same table.}
\label{tab:results}
\resizebox{0.4\textwidth}{!}{%
\begin{tabular}{|c|c|c|c|c|}
\hline
\multicolumn{5}{|c|}{\textit{Diagnosis from Chest X-Ray}}                                       \\ \hline
\multirow{2}{*}{Disease}          & \multirow{2}{*}{Metric} & \multicolumn{3}{c|}{Model comparison} \\ \cline{3-5} 
                                  &                         & Baseline       & Partial     & Full      \\ \hline
\multirow{3}{*}{Atelectasis}      & AUC                     & 0.865      & 0.870       & 0.873     \\ \cline{2-5} 
                                  & Precision               & 0.889      & 0.891       & 0.893     \\ \cline{2-5} 
                                  & Recall                  & 0.925      & 0.924       & 0.926     \\ \hline
\multirow{3}{*}{Edema}            & AUC                     & 0.898      & 0.883       & 0.884     \\ \cline{2-5} 
                                  & Precision               & 0.503      & 0.457       & 0.405     \\ \cline{2-5} 
                                  & Recall                  & 0.511      & 0.525       & 0.484     \\ \hline
\multirow{3}{*}{Pneumothorax}     & AUC                     & 0.829      & 0.837       & 0.857     \\ \cline{2-5} 
                                  & Precision               & 0.558      & 0.586       & 0.591     \\ \cline{2-5} 
                                  & Recall                  & 0.460      & 0.505       & 0.512     \\ \hline
\multirow{3}{*}{No Finding}       & AUC                     & 0.866      & 0.889       & 0.846     \\ \cline{2-5} 
                                  & Precision               & 0.346      & 0.336       & 0.349     \\ \cline{2-5} 
                                  & Recall                  & 0.188      & 0.313       & 0.313     \\ \hline
\multicolumn{5}{|c|}{\textit{Mammography Debiasing Results}}                                       \\ \hline
\multirow{2}{*}{Infer breast tissue densities} & \multirow{2}{*}{Metric} & \multicolumn{3}{c|}{Model comparison} \\ \cline{3-5} 
                                  &                         & Baseline       & Partial     & Full      \\ \hline
\multirow{3}{*}{1}                & AUC                     & 0.965      & 0.960       & 0.942     \\ \cline{2-5} 
                                  & Precision               & 0.637      & 0.709       & 0.637     \\ \cline{2-5} 
                                  & Recall                  & 0.858      & 0.677       & 0.682     \\ \hline
\multirow{3}{*}{2}                & AUC                     & 0.899      & 0.896       & 0.879     \\ \cline{2-5} 
                                  & Precision               & 0.781      & 0.769       & 0.763     \\ \cline{2-5} 
                                  & Recall                  & 0.765      & 0.758       & 0.736     \\ \hline
\multirow{3}{*}{3}                & AUC                     & 0.923      & 0.895       & 0.917     \\ \cline{2-5} 
                                  & Precision               & 0.879      & 0.825       & 0.831     \\ \cline{2-5} 
                                  & Recall                  & 0.739      & 0.727       & 0.821     \\ \hline
\multirow{3}{*}{4}                & AUC                     & 0.979      & 0.972       & 0.957     \\ \cline{2-5} 
                                  & Precision               & 0.413      & 0.324       & 0.482     \\ \cline{2-5} 
                                  & Recall                  & 0.867      & 0.883       & 0.625     \\ \hline
\end{tabular}%
}
\end{table}

\section{Conclusion}
We proposed a two step adversarial debiasing method with partial learning and evaluated the approach on two distinct medical image datasets. The proposed architecture can successfully preserved the targeted task performance while reducing the TPR disparity. This describes our initial experiments with the proposed methodology. In future, we plan to apply this technique for predictive analytics for healthcare problems while reducing the scocio-economical bias. The adversarial training approach described can be applied regardless of predictor’s model architecture, as long as the model is trained using a gradient-based method.

\newpage

\bibliography{ref} 
\bibliographystyle{aaai}
\end{document}